\documentclass{ws-procs9x6}
\usepackage{amsmath,amsfonts,ams,amssymb,wasysym}

\def\etal {{\it et al.}}

\begin{document}

\title{TERRESTRIAL VS. SPACEBORNE, QUANTUM VS. CLASSICAL TESTS OF THE EQUIVALENCE PRINCIPLE}
\author{M.\ HOHENSEE and H.\ M\"ULLER$^*$}

\address{Physics Department, University of California \\
Berkeley, CA 94720, USA\\
$^*$E-mail: hm@berkeley.edu}

\begin{abstract}
The equivalence principle can be tested by precision experiments based on classical and quantum systems, on the ground as well as in space. In many models, these tests are mostly equivalent in their ability to constrain physics beyond the Standard Model. We mention differences that nevertheless exist between spaceborne and quantum mechanical tests and their conventional competitors.
\end{abstract}

\bodymatter

\section{Introduction}

The Einstein equivalence principle (EEP) states that gravity is equivalent to acceleration, and affects all objects in exact proportion to their mass-energy. Experimental tests \cite{Nobili} of the EEP are often grouped into tests of Lorentz invariance \cite{datatables}, local position invariance \cite{LPI}, and the weak equivalence principle (WEP) \cite{WEP}, but this division is largely historical. Known `consistent' theories (e.g., energy and momentum-conserving) that violate one of these principles typically violate all. A conjecture attributed to Schiff \cite{Schiff} suggests that this must always be the case (although specialized counterexamples exist). Violations of the EEP are promising candidates for low-energy signals of Planck-scale physics \cite{Damour}.

The gravitational Standard-Model Extension (SME) \cite{SME} provides a starting point for this discussion, though we will also go beyond it. It is constructed from the lagrangians of the Standard Model and gravity by adding new interactions that violate Lorentz invariance and the EEP. For a fermion of mass $m$, for example, these interactions are encoded in eight Lorentz tensors $a_\mu, b_\mu, c_{\mu\nu}, d_{\mu\nu}, e_\nu, f_\nu, g_{\lambda\mu\nu},$ and $H_{\mu\nu}$ known collectively as coefficients for Lorentz violation. Leading-order violations of the EEP arise from the six $(a^{\rm p})_0, (a^{\rm n})_0, (a^{\rm e})_0, (c^{\rm p})_{00}, (c^{\rm n})_{00}, (c^{\rm e})_{00}$,
where the superscripts p, n, e denote the proton, the neutron, and the electron, respectively.
Experiments with neutral matter can only measure the combination $(a^{\rm p+e})_0 \equiv (a^{\rm p})_0+(a^{\rm e})_0$.
This leaves five measurable coefficients for EEP violations.\cite{Hohensee,SME}

We will not discuss technical aspects of space tests (e.g., interrogation time, short rotational and orbital periods, vibrations) and quantum tests (e.g., control of initial conditions and systematic effects, reproducibility, or the Coriolis force\cite{Coriolis}). A few unique signals for quantum tests have been discussed elsewhere, e.g., space-time foam \cite{Goklu} and dilatons \cite{Damour}.

\section{Unique signals for quantum tests}

In any relativistic framework, the phase of a matter wave is given by $\phi=-mc^2\tau/\hbar$, where $mc^2$ is the rest mass-energy and $\tau=\int d\tau$ the proper time along the path \cite{Dimopoulos,LPI,Hohensee}. Quantum tests measure the difference of that phase between two paths. In General Relativity (GR) and theories that satisfy Schiff's conjecture (e.g., the SME), changes in the proper time $\tau$, which would be measurable by atom interferometers \cite{LPI,Hohensee} and clocks, will lead to changes in the center of mass motion, which would be measurable by classical WEP tests. 
However, in theories that go beyond Schiff's conjecture, signals such as nonstandard gravitational redshifts might be picked up by clocks and quantum WEP tests even if they are undetectable by classical WEP tests, giving these tests a distinct power.

Classical and quantum tests differ in the species they use. 
Quantum tests typically use atoms with a simple electronic structure, e.g., alkalis. They are not suitable for use in classical tests, as they are soft and chemically reactive, but have a special nuclear structure. Use of such atoms is important for measuring all types of EEP violations \cite{Hohensee}. Use of (anti-)hydrogen in quantum experiments or hydrogen-rich materials in classical ones  would be of particular interest.

Spin-dependent gravitational couplings have long been studied in the context of theories of gravity with nonvanishing torsion \cite{Spin}. In the SME, such effects are expected to result from the $b$, $d$, $g$, and $H$ coefficients. The the $e$-type coefficients are not included here, as they can be lumped into the $a$-type coefficients. Similarly, the $f$-type can be lumped into the $c$-type coefficients.\cite{f-coeff} These coefficients will describe how gravity might couple differently to particles exhibiting different spin-orbit couplings, i.e., having correlated external and spin degrees of freedom. 
Only quantum experiments will be able to study them. 
For example, an experiment reaching $10^{-14}$ or better could perform the measurement of several components of $b^{\rm p}, d^{\rm p}, g^{\rm e}, g^{\rm n}, g^{\rm p},$ and $H^{\rm p}$, for which limits do not exist at the time of this writing.  The theory of these effects might start from the known relativistic hamiltonian for the gravitational SME, deriving the nonrelativistic hamiltonian by a Foldy-Wouthuysen transformation \cite{SME}. Most of the additional coefficients should lead directly to observable effects. 
The fluctuations with gravity might make additional ones observable.

\section{Unique signals for spaceborne tests}

Letting an EEP-test experiment perform measurements over a year and analyzing its signal for periodic variations at the frequencies of Earth's rotation and orbit and their combinations will allow us to separately search for effects caused by the Earth's and the Sun's gravity. Spaceborne experiments offer an additional way for such separate measurement through their orbit. Different SME parameters describing the source mass of the gravitational field can thus be constrained. Since the combinations of coefficients that produce shifts in active and passive gravitational masses are not the same \cite{SME}, and because of the different composition of the Earth and the Sun, this may result in  sensitivity to additional parameter combinations. Explicit calculation has not been performed yet.

Effects of GR can be ordered by powers of the dimensionless gravitational potential $U/c^2$ and particle velocity $(v/c)^2$. 
Nontrivial higher order effects arise in GR due to the theory's nonlinearity, e.g., perihelion precession. In atom interferometry, they generate extra phase shifts\cite{Dimopoulos} proportional to $(U/c^2)(v/c)^2$.

Higher-order SME terms in flat spacetime \cite{KosteleckyMewes2012} are likely to cause EEP-violations at $O(1/c^3)$ and higher, though this has not been studied yet. Such effects will go beyond the traditional categorization into tests of Lorentz invariance, WEP, and local position invariance: The validity of EEP for particles at rest relative to the source mass might not imply its validity for moving objects, or its validity at one location might not imply its validity elsewhere. These effects should be studied further. 
An important figure of merit for experiments searching for high-order effects is the velocity the experiments attain with respect to the source mass. Spaceborne experiments clearly outperform laboratory experiments in this regard.

A chameleon is a hypothetical scalar field $\phi$ proposed to help explaining dark matter and the accelerated expansion of the universe.\cite{Chameleon} Through nonlinear self-coupling the chameleon is a short (millimeter)-range force close to massive objects, such as Earth. In empty space, however, it may give rise to a long-range fifth force that causes order-unity equivalence principle violations. Chameleons thus avoid detection in laboratory and solar system experiments, but might cause large EEP violations for small test particles in empty space. While a detailed analysis remains to be performed, it seems likely that models that are compatible with all previous tests could still lead to EEP violations between $10^{-19}$ and $10^{-11}$ .

Finally, we might speculate that any hints at distance-dependent gravity anomalies (perhaps varying Hubble constant \cite{Hubble}) are a reason to search for EEP violations at all accessible distance scales, by Earth-and space-based experiments, solar system tests, as well as astrophysics.

\section*{Acknowledgments}
We thank Alan Kosteleck\'y, Anna Nobili, Jim Phillips, Ernst Rasel, Andreas Sch\"arer, and Nan Yu for important discussions and the David and Lucile Packard Foundation, the National Aeronautics and Space Agency, and the National Science Foundation for support.

\end{document}